\documentclass[onecolumn,11pt,a4paper,aps]{revtex4}

\newcommand{\eq}[1]{\begin{equation}#1\end{equation}}
\newcommand{\dd}{\mathrm{d}}
\newcommand{\ee}{\mathrm{e}}

\newcommand{\ch}{\mathrm{ch\,}}
\newcommand{\sh}{\mathrm{sh\,}}

\usepackage{amsmath}
\usepackage{graphics}
\usepackage{graphicx}
\usepackage{psfrag}

\begin{document}

\title{Exact results for the entanglement across defects in
critical chains}
\author{Ingo Peschel$^1$ and Viktor Eisler$^2$}
\affiliation{
$^1$Fachbereich Physik, Freie Universit\"at Berlin,
Arnimallee 14, D-14195 Berlin, Germany\\
$^2$Vienna Center for Quantum Science and Technology, Faculty of Physics,\\
University of Vienna, Boltzmanngasse 5, A-1090 Vienna, Austria
}

\begin{abstract}
We consider fermionic and bosonic quantum chains where a defect separates two subsystems
and compare the corresponding entanglement spectra. With these, we calculate their R\'enyi 
entanglement entropies and obtain analytical formulae for the continuously varying coefficient 
of the leading logarithmic term. For the bosonic case we also present numerical results.
\end{abstract}

\maketitle

\section{Introduction}

  In critical quantum chains, the entanglement entropy between a section of length $L$ 
and the remainder varies as $\ln L$ with a prefactor proportional to the central charge $c$
of the model. For a review, see \cite{CC09}. If one modifies the interface,
this coefficient has been found to vary continuously with the defect strength
in free particle systems. The effective number of states in the Schmidt 
decomposition then increases as a power of $L$ with a continuously varying exponent. 

For fermionic systems, the problem was first posed by Levine 
\cite{Levine04} and then investigated numerically for XX chains \cite{Peschel05} and transverse 
Ising chains \cite{Igloi/Szatmari/Lin09}. By mapping the problem to that of a two-dimensional 
Ising model with a defect line, the coefficient could later be obtained analytically 
and perfect agreement with the numerical data was found \cite{Eisler/Peschel10}. Moreover, 
the parameter entering the analytical expression turned out to be simply the transmission 
amplitude through the defect. This holds also for more complicated defects
\cite{Eisler/Garmon10}. In a series of recent papers, calculations were also done for continuous 
fermionic systems, and the same coefficient was found \cite{CMV11a,CMV11b,CMV11c}. 
On the bosonic side, Sakai and Satoh studied a continuum system with a conformal interface 
between two different critical parts with $c=1$ \cite{Sakai/Satoh08}. This can also be viewed 
as a uniform system with a defect. The continuously varying coefficient found in this case 
has a close relation to the fermionic one, but the detailed connection has remained unexplored 
so far.
 
  The purpose of this note is two-fold. Firstly, we want to show how one can treat the bosonic
case in complete analogy to the fermionic one. Thus we consider a system of coupled oscillators 
which is the lattice version of the system studied in \cite{Sakai/Satoh08}. For this system, we
derive an expression for the single-particle spectrum in the reduced density matrix $\rho$ which is
formally very similar to that in the fermionic case. However, it does not have a gap, since
the defect does not break the criticality. This clarifies the relation between the two problems
at the level of the RDM spectra. We check that one recovers the previous result for the von Neumann 
entropy in this way and also compare with numerics.

  Secondly, with the spectra at hand, we show that not only the von Neumann entropy 
$S= -\mathrm{tr}\,(\rho \ln \rho)$, but also the R\'enyi entropies 
\begin{equation}
S_{n}= \frac {1}{1-n} \mathrm{ln\,tr}(\rho^n)
\label{renyi_def}
\end{equation}
can be calculated asymptotically in closed form for all integer $n$. In both cases, the coefficients 
$\kappa_n$ in
\begin{equation}
S_{n}= \kappa_n \ln L 
\label{kappa_def}
\end{equation}
turn out to be (sums of) elementary functions for $n>1$ and in this sense simpler than the von 
Neumann coefficients. Qualitatively, they all vary in a similar way for each of the two cases.

In the following, we first recapitulate the fermionic results in section 2. Then, in section 3,
we study the oscillator chain and its RDM spectrum. In section 4 and 5 we present the calculations
for the R\'enyi entropies and show the resulting functions. In section 6 we sum up our findings
and in the appendix we present a derivation of the bosonic spectrum from the transfer matrix of
a two-dimensional Gaussian model.

\section{Setting and fermionic results}

We consider open chains of length $2L$ with a defect in the middle and the entanglement 
between left and right halves. The geometry is shown in fig. 1 for a bond defect.

%
%
\begin{figure}[htb]
\centering
\includegraphics[scale=0.5]{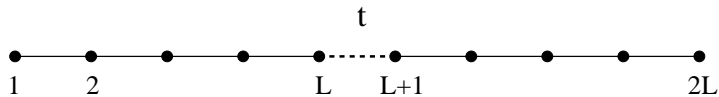}
\caption{Quantum chain with a bond defect.}
\label{fig:bond_defect}
\end{figure}
%
For free-particle systems, the reduced density matrix for a subsystem can be written 
\begin{equation}
\rho = \frac{1}{Z}\; e^{-\mathcal{H}}
\label{rhogen}
\end{equation}
where $\mathcal{H}$ is again a free-particle Hamiltonian, see \cite{review09}.
Its single-particle eigenvalues, called $2\omega_k$ in the following,
contain the basic entanglement information. With them, the von Neumann entropy $S$ is given by
\begin{equation}
S = \pm \sum_k \ln(1 \pm \mathrm{e}^{-2\omega_k})
       +\sum_k \frac {2\omega_k}{\mathrm{e}^{2\omega_k} \pm 1}
\label{neumann_gen}
\end{equation}
and the R\'enyi entropies are
\begin{equation}
S_{n}= \frac {1}{1-n} \big[ \pm \sum_k \ln(1 \pm \mathrm{e}^{-2n\omega_k})
       \mp n \sum_k \ln(1 \pm \mathrm{e}^{-2\omega_k})\big]
\label{renyi_gen}
\end{equation}
where the upper (lower) sign refers to fermions (bosons).

For a transverse Ising model, these eigenvalues were determined in \cite{Eisler/Peschel10} 
from the excitations in the transfer matrix of an Ising strip of width $\ln L$. The result 
was that 
\eq{
\ch \omega_k =  \frac{1}{s}\,\ch \varepsilon_k
\label{omega_ferm}}
Here the $\varepsilon_k$ are the values without the defect and vary linearly with $k$ for
$\ln L \gg 1$, while $s$ measures the defect strength. For a bond defect, where the coupling 
is changed from 1 to $t$, it is given by $s = \sin(2\arctan t) =2/(t+1/t)$. For an XX chain,
$T = s^2$ is the transmission coefficient through the defect at the Fermi level. The dispersion 
relation (\ref{omega_ferm}), shown in fig. 5 of \cite{Eisler/Peschel10}, describes a spectrum 
with a gap induced by the defect and encodes the entanglement properties for a large system. 
In particular, the entanglement decreases as $s$ becomes smaller. This was investigated in 
\cite{Eisler/Peschel10} for the von Neumann entropy.

The functional relation (\ref{omega_ferm}) can also be written in another form. In terms of the 
quantities with and without the defect, 
\begin{equation}
\zeta_k'=1/(e^{2\omega_k}+1) , \quad \quad \zeta_k=1/(e^{2\varepsilon_k}+1)
\label{occupation_fermi}
\end{equation}
it becomes
\begin{equation}
\zeta_k'(1-\zeta_k') = s^2 \zeta_k(1-\zeta_k)
\label{functional_fermi}
\end{equation}
Such a relation was found recently in the study of quantum wires, i.e. for continuous systems with 
a localized, scale-free scattering potential \cite{CMV11b,CMV11c}. In these calculations, one 
considers the overlap matrix $\bf{A}$ of the occupied single-particle states in the subsystem 
\cite{Klich06} and finds that
\begin{equation}
{\bf{A'(1-A')}} = s^2 {\bf{A(1-A)}}
\label{functional_overlap}
\end{equation}
for arbitrary particle number $N$. This relation extends to the eigenvalues $a_k$ which 
are then used to determine the entanglement entropies.  However, for free particles the 
non-trivial $a_k$ are the same as the eigenvalues $\zeta_k$ of the correlation matrix $\bf{C}$ which
give the $\omega_k$ and $\varepsilon_k$ via (\ref{occupation_fermi}). The relation for $a_k$ in 
\cite{CMV11b,CMV11c} is therefore the same as (\ref{functional_fermi}) and the calculations, 
although they proceed in a different way, have actually the same basis and thus lead to the 
same results.

\section{Bosonic chain}

A quantum chain realizing the system studied in \cite{Sakai/Satoh08} consists of $2L$ harmonic
oscillators coupled by springs, where the spring constants and the masses are different 
in both halves, but have the same ratio, see \cite{Bachas07}. The spring in the centre has to 
be chosen properly.
The Hamiltonian is
\begin{equation}
H= \sum_{n=1}^{2L} \left(- \frac {1}{2m_n} \frac {\partial^2}{\partial x^2_n}+
\frac {1}{2} m_n \Omega_0^2\, x^2_n  \right) + \frac {1}{2} \sum_{n=1}^{2L-1} D_n (x_n- x_{n+1})^2
\label{H_osc}
\end{equation}
and we set 
\begin{equation}
D_n=m_n = \left\{
\begin{array}{ll}
K_1= \ee^{\theta} & \quad n < L \\
K_2= \ee^{-\theta} & \quad n > L
\label{parameters_osc1}
\end{array}\right.
\end{equation}
while the central spring is assumed to have
\begin{equation}
D_L = K_0 \equiv \frac {2K_1K_2}{K_1+K_2} = \frac {1}{\ch \theta}
\label{parameters_osc2}
\end{equation}
A rescaling of the coordinates $u_n=\sqrt{m_n}\, x_n$ then makes $H$ homogeneous up to the 
springs at sites $L$ and $L+1$
\begin{equation}
H= \sum_{n=1}^{2L} \left(- \frac {1}{2} \frac {\partial^2}{\partial u^2_n}+
\frac {1}{2} \Omega_0^2\, u^2_n  \right) + \frac {1}{2} \sum_{n \neq L} (u_n- u_{n+1})^2
+\frac {1}{2}\left( \frac {K_0}{K_1} u_L^2 + \frac {K_0}{K_2} u_{L+1}^2 
- 2 \frac {K_0}{\sqrt{K_1K_2}} u_Lu_{L+1} \right)
\label{H_osc}
\end{equation}
In this form, one is dealing with a defect problem, and the defect is completely
characterized by the parameter $\theta$. For large $|\theta|$, i.e. if $K_1$ and $K_2$ are
very different, the chain is cut in the middle.
The eigenfrequency $\Omega_0$ of the
oscillators is included to avoid a zero mode, but will be taken small.

The single-particle eigenvalues in the RDM follow from the matrix ${\bf{C}}=2{\bf{X}}\,{2\bf{P}}$
containing the position and momentum correlations in the subsystem \cite{review09}. In terms of the
eigenvalues $\Omega_m^2$ and eigenfunctions $\phi_m(i)$ of the dynamical matrix, one has
\begin{equation}
C_{ij} = \sum_{m,n=0}^{2L-1}\frac{\Omega_n}{\Omega_m}A_{mn} \phi_m(i)\phi_n(j)
\label{eq:cij1}
\end{equation}
where the reduced overlap matrix 
\begin{equation}
A_{mn}=\sum_{l=1}^{L} \phi_m(l)\phi_n(l)
\label{overlap1}
\end{equation}
comes from taking the product of ${\bf{X}}$ and ${\bf{P}}$ in the subsystem (chosen as the 
left half-chain). 

In the $\it{homogeneous}$ system
\eq{
\Omega_m = \sqrt{\Omega_0^2+2\left(1-\cos \frac{m \pi} {2L}\right)}, \quad m = 0,\dots, 2L-1
}
and 
\begin{equation}
\phi_m(i)=\sqrt{\frac{1}{L}} \cos \frac{\left(i- 1/2\right) m \pi}{2L} \quad m \ne 0, \quad
\phi_0(i) = \sqrt{\frac{1}{2L}}
\label{eigenfunctions_hom}
\end{equation}
Then the matrix $A$ is for $m,n \ne 0$
\begin{equation}
A_{mn} = \frac{1}{4L} \left[ 
\frac{\sin\frac{\pi}{2}(m-n)}{\sin\frac{\pi}{4L}(m-n)} +
\frac{\sin\frac{\pi}{2}(m+n)}{\sin\frac{\pi}{4L}(m+n)} 
\right]
\label{overlap2}
\end{equation}
and if one or both the indices are zero one has
\begin{equation}
A_{m0} = A_{0m} =\frac{1}{\sqrt{2} 2L}
\frac{\sin\frac{\pi}{2}m}{\sin\frac{\pi}{4L}m}  , \quad
A_{00} = 1/2
\label{overlap3}
\end{equation}
Note, that $A_{mn}$ vanishes if $m-n \ne 0$ is even. Furthermore, $A_{mm}=1/2$ and one can write
\eq{
C_{ij} = \frac{1}{2} \delta_{ij} + D_{ij}
\label{eq:cij2}}
where $D_{ij}$ is the piece of (\ref{eq:cij1}) with the sum restricted to $m-n$ odd.

For the $\it{inhomogeneous}$ system, the eigenvalue equations can be satisfied by choosing 
the same spectrum as in the homogeneous case $\Omega'_m = \Omega_m$ and making the ansatz 
\eq{
\phi'_m(i) = \left\{
\begin{array}{ll}
\alpha_m \phi_m(i) & \quad 1 \le i \le L \\
\beta_m \phi_m(i) & \quad L <  i \le 2L
\end{array}\right.
\label{eigenfunctions_inhom}
}
Inserting this into the two modified equations and requiring orthonormality, one obtains the 
conditions
\eq{
\alpha_m \beta_m = K_0, \quad \alpha_m \alpha_n + \beta_m \beta_n = 2 \delta_{mn}
}
The solutions are
\eq{
\alpha_m^2 = 1 \pm \tanh \theta, \quad \beta_m^2 = 1 \mp \tanh \theta
}
where the upper (lower) signs refer to even (odd) indices $m$. This yields
\eq{
A'_{mn} = \left\{
\begin{array}{lll}
A_{mn} (1-\tanh^2\theta) & \mbox{$m-n$ odd}\\
A_{mn} (1+\tanh \theta)^2 &  \mbox{$m=n$ even} \\
A_{mn} (1-\tanh \theta)^2 &  \mbox{$m=n$ odd}
\end{array}\right.
}
Finally, one obtains
%
\begin{equation}
\begin{split}
C'_{ij} & = \frac{1}{2} (1 + \tanh^2 \theta) \delta_{ij} + (1-\tanh^2\theta)D_{ij} \\
        & = (1-\tanh^2\theta) C_{ij} + \tanh^2 \theta \,\delta_{ij} 
\end{split}
\label{functional_matr}
\end{equation}
%
which is an exact relation between the matrices with and without the defect and translates
to their eigenvalues $\coth^2 \omega_k$ and $\coth^2 \varepsilon_k$. Written differently, it 
takes the form
\eq{
\sh \omega_k =  \frac{1}{s}\,\sh \varepsilon_k 
\label{omega_bos}}
where $s = 1/\ch \theta$. This is the analogue of the relation (\ref{omega_ferm}) and has 
a striking similarity to it. However, while the two relations are identical for large $\omega$ and 
$\varepsilon$, the lower part of the spectrum is different. There is no gap in (\ref{omega_bos}),
$\omega$ approaches zero with slope $1/s$ as $\varepsilon$ goes to zero. The defect does not make 
the system non-critical. Full dispersion curves are shown in fig. 2 for several $s$. Also marked
are the discrete numerical values which are obtained for a chain of $L=200$ sites. We also
note that with 
\begin{equation}
n_k'=1/(e^{2\omega_k}-1) , \quad \quad n_k=1/(e^{2\varepsilon_k}-1)
\label{occupation_bose}
\end{equation}
the functional equation can be written as
\begin{equation}
n_k'(1+n_k') = s^2 n_k(1+n_k)
\label{functional_bose}
\end{equation}
which is the analogue of (\ref{functional_fermi}). Furthermore, one should mention that the overlap
matrix (\ref{overlap2}) has a similar structure as the one in \cite{CMV11b,CMV11c} and 
also satisfies (\ref{functional_overlap}). The two calculations are therefore closely related.

As in the fermionic case, the parameter $s$ has the meaning of a transmission amplitude through
the defect. This can be seen from the $2 \times 2$ transfer matrix for the scattering problem.
This matrix has eigenvalues ($\ee^{\theta}, \ee^{-\theta}$) and transmits the odd resp. even functions 
(\ref{eigenfunctions_hom}) by multiplying them with these factors. This explains the form 
(\ref{eigenfunctions_inhom}) of the perturbed eigenfunctions. One can also see, that only the
choice (\ref{parameters_osc2}) leads to a transmission coefficient $T=s^2$ independent of the
wavelength (i.e. to a scale-free defect) and to such a simple structure of the problem. 

The relation (\ref{omega_bos}) can also be derived by going to two dimensions and studying
the transfer matrix of a Gaussian model. This is sketched in the Appendix. The relation is also
%
%
\begin{figure}
\center
\includegraphics[scale=0.7]{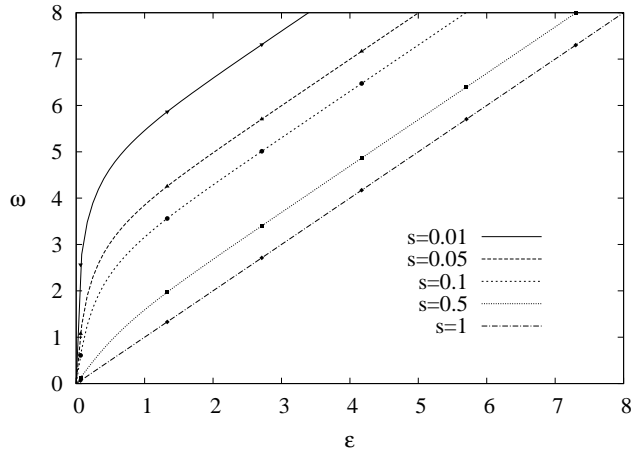}
\caption{The relation (\ref{omega_bos}) for the bosonic single-particle excitations in 
$\mathcal{H}$ for several values of the defect strength. The points are the numerical values 
for chains with $L=200$.}
\label{fig:omega_bos}
\end{figure}
%
%
implicit in the work of \cite{Sakai/Satoh08}, but not really visible.
However, one can insert it into the bosonic von Neumann entropy (\ref{neumann_gen}),
change to integrals and obtain $\kappa=I/\pi^2$ with
\begin{equation}
I = -\int_{0}^{\infty} \dd \varepsilon \;\ln(1-\mathrm{e}^{-2\omega_k})
       +\int_{0}^{\infty} \dd \varepsilon \; \frac {2\omega_k}{\mathrm{e}^{2\omega_k}-1}
\label{integral_bos}
\end{equation}
Expressing this in terms of hyperbolic functions, differentiating twice with respect to $s$ and
using a partial integration gives, with $x=\sh \varepsilon$
\begin{eqnarray}
I''(s) &=& \int_{0}^{\infty} \dd x \frac{x}{\sqrt{x^2+1}^3 \sqrt{x^2+s^2}} \mathrm{arsh}(\frac {x}{s})
            \nonumber \\
        &=& \frac{1}{1-s^2} \ln s 
\label{secderiv_I}
\end{eqnarray}
This is (4.12) in \cite{Sakai/Satoh08} and the negative of the corresponding result
in the fermionic case \cite{Eisler/Peschel10}. A further difference appears in the integration
over $s$, since $I'(0)=\pi^2/4$ here, while it vanishes for fermions. Therefore $\kappa(s)$
contains a term linear in $s$ and reads
\begin{equation}
\kappa(s) = \frac {1}{4} s - \kappa_F(s)
\label{kappa_bos}
\end{equation}
where the second part is the fermionic result 
\eq{
\kappa_F(s)=-\frac{1}{2\pi^2} \left\{ \left[ (1+s) \ln (1+s) + (1-s) \ln (1-s) \right] \ln s
+ \left[ (1+s) \mathrm{Li_2}(-s) + (1-s) \mathrm{Li_2}(s)\right] \right\}
\label{eq:intint}}
with $\mathrm{Li_2}$ denoting the dilogarithm. This is, written somewhat differently, the final result in 
\cite{Sakai/Satoh08}. The function $c_\mathrm{eff}=6\kappa(s)$ is shown in fig. 3 and rises smoothly 
from zero to 1, while $\kappa(s)$ itself varies between 0 and 1/6. 
%
%
\begin{figure}
\center
\includegraphics[scale=0.7]{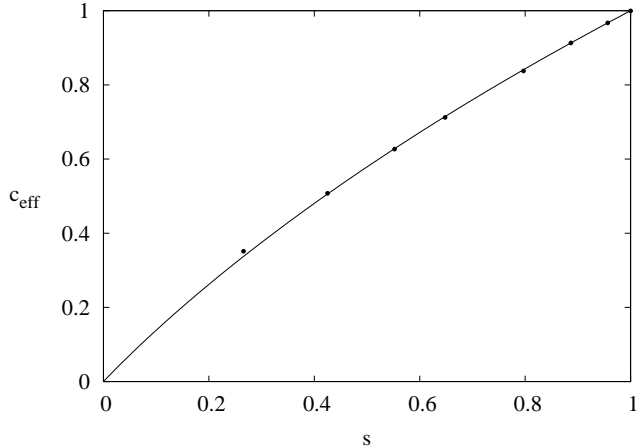}
\caption{The function $c_\mathrm{eff}=6\kappa(s)$ for the bosonic interface problem. The points are
numerical results, see text.}
\label{fig:kappa_bos}
\end{figure}
%
%

The figure also contains data points from numerical calculations. In order to see a logarithmic
behaviour of $S$ in the bosonic case, one has to work in a situation, where the correlation length
given by $1/\Omega_0$ is larger than the system size but still finite. In the homogeneous system
one can then obtain $\kappa=1/6$ with high precision (3-4 decimal places) from sizes $L=100-500$.
The calculations are done such that $2L\,\Omega_0=A$ is held constant as $L$ is varied. 
With the defect, the dispersion curve shown in fig. \ref{fig:omega_bos} becomes rather steep for 
small $s$ and only a single eigenvalue is found on the slope for usual sizes $2L$. The results
depend sensitively on this eigenvalue and deviate from the theoretical curve for small $s$. 
We have therefore plotted only data for $s > 1/4$, obtained with $A=0.01$, where the agreement is 
very good.

\section{Fermionic R\'enyi entropy}

We now turn to the R\'enyi entropy. In terms of the single-particle eigenvalues, it is
given by (\ref{renyi_gen}).
Converting the sum over $k$ into an integral over $\varepsilon$ leads to the logarithmic
behaviour (\ref{kappa_def}) of $S_n$ with coefficient 
\begin{equation}
\kappa_{n}= \frac {1}{\pi^2} \, \frac {1}{1-n} K_n\, , \quad \quad K_n = I_n - n I_1
\label{kappa_gen}
\end{equation}
where
\begin{equation}
I_n = \int_{0}^{\infty} \dd \varepsilon \; \ln(1+ \ee^{-2n\omega}) 
\label{integral_I_n}
\end{equation}
In terms of hyperbolic functions, the quantity $K_n$ is
\begin{equation}
K_n = \int_{0}^{\infty} \dd \varepsilon \; \left[ \ln(2 \ch n\omega)- n \ln(2 \ch \omega) \right]
\label{integral_K_n}
\end{equation}
To evaluate it, one first writes $\ch n\omega$ as a product by using 1.391 of \cite{GR65}. For even
$n$, this gives
\begin{equation}
\ch n\omega = \prod_{k=1}^{n/2} \left[\frac{\ch^2 \omega -c_k^2}{s_k^2} \right]  
\label{product_even}
\end{equation}
where $s_k=\sin((2k-1)\pi/2n)$ and $c_k=\cos((2k-1)\pi/2n)$. One now inserts $\ch \omega = 
\ch \varepsilon/s$ and takes the derivative with respect to $s$. This gives, with
$x=\ch \varepsilon$,
\begin{equation}
\frac{\partial}{\partial s}\ln (2\ch n\omega) = 
                                - \frac{n}{s}- \sum_{k=1}^{n/2} \frac{2sc_k^2}{x^2-s^2c_k^2}
\label{deriv_lncosh}
\end{equation}
The first term is compensated by an identical one from $n \ln(2 \ch \omega)$ and one has
\begin{eqnarray}
K_n'(s) &=&  \sum_{k=1}^{n/2} c_k  \int_{1}^{\infty} \frac{\dd x}{\sqrt{x^2-1}} \left[\frac{1}{x-sc_k}
           -\frac{1}{x+sc_k} \right] \nonumber \\
        &=&  \sum_{k=1}^{n/2} c_k \frac{2 \arcsin(sc_k)}{\sqrt{1-s^2c_k^2}} 
\label{deriv_K_n}
\end{eqnarray}
which can be integrated to give
\begin{equation}
K_n(s) = -  \sum_{k=1}^{n/2}  \arcsin^2(sc_k)
\label{result_K_even}
\end{equation}
With $k \rightarrow (n/2+1-k)$ the $c_k$ can be changed to $s_k$ and one finally obtains
\begin{equation}
\kappa_n(s) = \frac {1}{\pi^2} \, \frac {1}{n-1} \sum_{k=1}^{n/2}  \arcsin^2(ss_k), \quad
               s_k=\sin(\frac{(2k-1)\pi}{2n}), \quad n \,\,\, even
\label{kappa_even}
\end{equation}
For odd $n$, the expression (\ref{product_even}) is changed into
\begin{equation}
\ch n\omega = \ch \omega \prod_{k=1}^{(n-1)/2} \left[\frac{\ch^2 \omega -c_k^2}{s_k^2} \right]  
\label{product_odd}
\end{equation}
but leads to the same result as in (\ref{result_K_even}), up to the summation limit. Replacing
the $c_k$ again by sine functions, one has  
\begin{equation}
\kappa_n(s) = \frac {1}{\pi^2} \, \frac {1}{n-1} \sum_{k=1}^{(n-1)/2}  \arcsin^2(s\bar s_k), \quad
               \bar s_k=\sin(\frac{2k\pi}{2n}), \quad n \,\,\, odd
\label{kappa_odd}
\end{equation}

\noindent Formulae (\ref{kappa_even})and (\ref{kappa_odd}) give the $\kappa_n$ for all integer $n$. 
One sees, that they are all elementary functions, namely sums of $n/2$ or $(n-1)/2$ $\arcsin^2$ 
terms. This is in contrast to the limiting cases $n \rightarrow 1$ (von Neumann entropy) and 
$n \rightarrow \infty$ (largest eigenvalue of $\rho$), where the dilogarithm $\mathrm{Li_2}$ appears 
\cite{Eisler/Peschel10}. The result is particularly simple for $n=2,3$ where only a single term
is present
\begin{eqnarray}
\kappa_2(s) &=&  \frac{1}{\pi^2} \arcsin^2(s/\sqrt{2}) \\
\kappa_3(s) &=&  \frac{1}{2\pi^2} \arcsin^2(s\sqrt{3}/2)
\label{kappa_23}
\end{eqnarray}
The first formula was given before in \cite{Peschel11}.
%
%
\begin{figure}
\center
\includegraphics[scale=0.9]{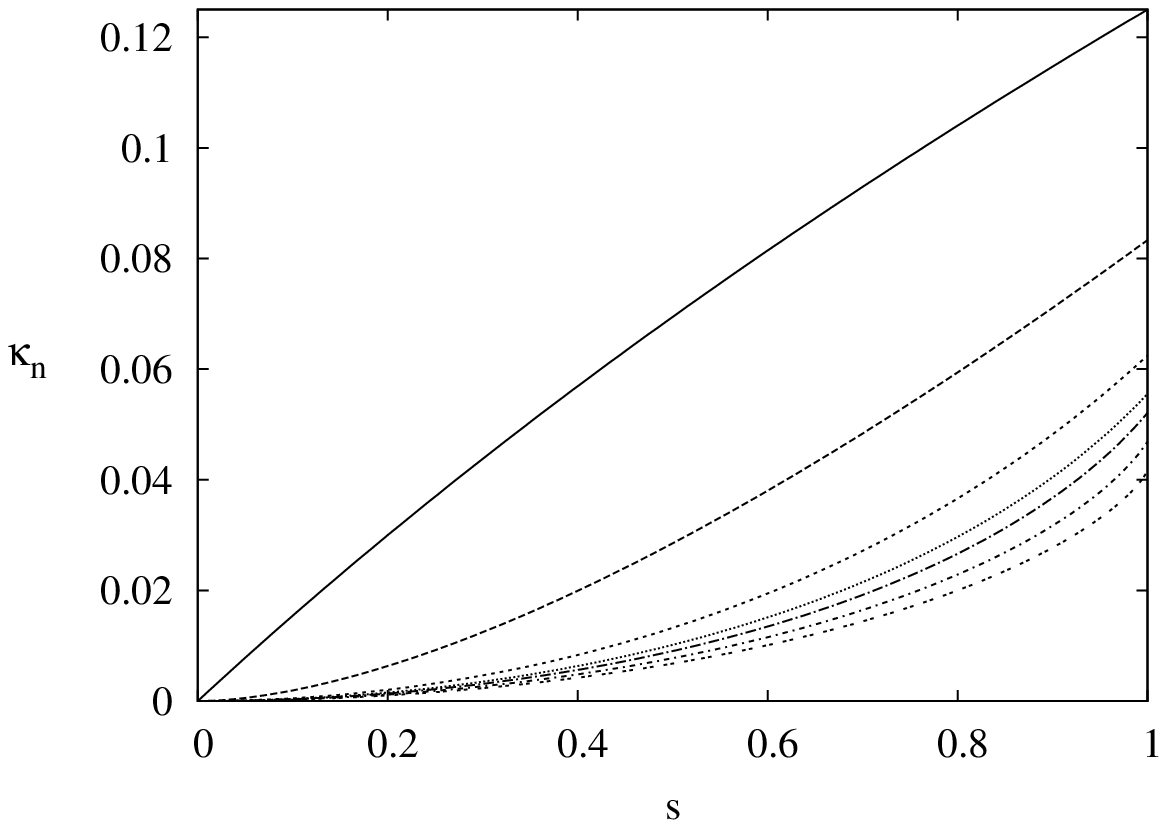}
\includegraphics[scale=0.9]{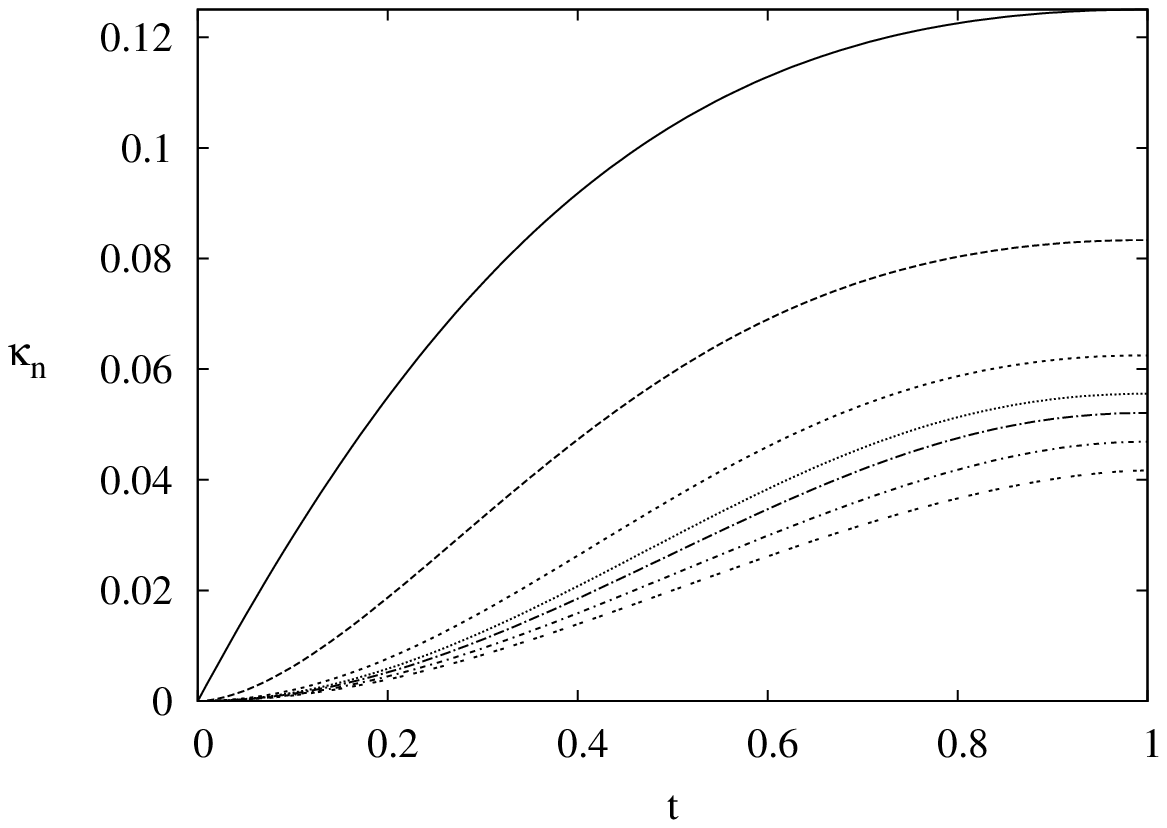}
\caption{Coefficients $\kappa_n$ for a transverse Ising chain with a defect.
Upper panel: as a function of the parameter $s$. Lower panel: as a function of the defect
bond strength $t$. The $n$-values are, from top to bottom, $n=1/2,1,2,3,4,8,\infty.$}
\label{fig:kappa_fermion}
\end{figure}
%
%

Qualitatively, the $\kappa_n$ all vary similarly with $s$, rising from zero for $s=0$ (the dissected
system) to a limiting value
\begin{equation}
\kappa_n(1) = \frac {1}{24} (1+ \frac {1}{n})
\label{kappa(1)}
\end{equation}
for $s=1$ (the homogeneous system) which can be obtained either directly from $I_n(1)= \pi^2/24n$,
or by carrying out the simple sums which remain in $\kappa_n$ for $s=1$. For $n=1$, 
$\kappa_1(1)=1/12=c/6$ with $c=1/2$. The behaviour near $s=0$ is quadratic and given by
\begin{equation}
\kappa_n(s) = \frac {1}{4 \pi^2} \frac {n}{n-1}\, s^2 \, ,\quad s \rightarrow 0
\label{kappa(0)}
\end{equation}
As $n \rightarrow 1$, the curvature diverges which signals the $s^2\ln(1/s)$-behaviour one has in
the von Neumann entropy.

The $\kappa_n(s)$ are shown for several values of $n$ in the upper part of fig. 4. In the lower part,
the same $\kappa_n$ are plotted as functions of the bond variable $t$. The linear behaviour
near $s=1$ is then turned into a quadratic one near $t=1$, and the functions become symmetric under
$t \rightarrow 1/t$. 

Fig. 4 includes the results for $n=1$ and for $n=\infty$ which were found in \cite{Eisler/Peschel10}.
Also shown is the quantity $\kappa_n$ for $n=1/2$, which can be obtained from the derivative
\begin{eqnarray}
K_{1/2}'(s) &=&  \frac {1}{2} \int_{1}^{\infty} \frac{\dd x}{\sqrt{x^2-1}} \frac{1}{x+s} \nonumber \\
        &=&  \frac {1}{2} \frac{1}{\sqrt{1-s^2}}(\pi-\arcsin{s}) 
\label{deriv_K_1/2}
\end{eqnarray}
which gives
\begin{equation}
\kappa_{1/2}(s) = \frac {1}{2\pi^2} \, \arcsin s \, (\pi-\arcsin s)
\label{kappa_1/2}
\end{equation}
It varies linearly, $\kappa_{1/2}(s) \simeq s/2\pi$, for small $s$, has negative curvature everywhere
and lies significantly above the $n=1$ curve, but otherwise fits into the overall scheme.

The formulae given above all refer to the transverse Ising model, where $c=1/2$. For the XX (hopping)
model with $c=1$, they have to be multiplied by a factor of two. In this case, it is also easy to 
construct a scale-free defect in analogy to the bosonic chain. One only has to supplement the modified 
bond $t$ with site energies $\pm \sqrt{1-t^2}$ at $L$ and $L+1$, respectively. Then 
(\ref{functional_overlap}) is satisfied, the transmission amplitude is $s=t$ and the $\kappa_n(s)$ 
are the relevant quantities.

These results can also be obtained for continuum systems by working with the overlap matrices.
Then the particle number $N$ appears in (\ref{kappa_def}) instead of $L$ and the $\kappa_n$ are 
found in the form of an infinite series in the parameter $s$ which can be recognized as that of 
the function $\mathrm{arcsin}^2$ \cite{CMV11b,CMV11c}. In the cited papers, the notation is somewhat
different: $2\kappa_n$ is called ${\cal{C}}^{(n)}$ and $T$ is used instead of $s$.

\section{Bosonic R\'enyi entropy}

The calculation of $S_n$ in the bosonic case is very similar. The quantity $K_n$ introduced in 
(\ref{kappa_gen}) becomes
\begin{equation}
K_n = -\int_{0}^{\infty} \dd \varepsilon \; \left[ \ln(2 \sh n\omega)- n \ln(2 \sh \omega) \right]
\label{integral_K_n_bos}
\end{equation}
and the necessary formula for $\sh n\omega$ is, for odd $n$
\begin{equation}
\sh n\omega = n \; \sh \omega \prod_{k=1}^{(n-1)/2} \left[1+\frac{\sh^2 \omega}{\bar s_k^2} \right] 
              \, , \quad \bar s_k=\sin(k\pi/n) 
\label{product_odd_sh}
\end{equation}
This gives the derivative 
\begin{eqnarray}
K_n'(s) &=&  -\sum_{k=1}^{(n-1)/2} 2s\bar s_k^2\int_{0}^{\infty} \frac{\dd x}{\sqrt{x^2+1}} 
               \frac{1}{x^2+s^2\bar s_k^2} \nonumber \\
        &=&  -\sum_{k=1}^{(n-1)/2} \frac{2\bar s_k}{\sqrt{1-s^2\bar s_k^2}} 
               \arcsin{(\sqrt{1-s^2\bar s_k^2})}
\label{deriv_K_n_bos}
\end{eqnarray}
and the integrations leads to
\begin{equation}
K_n(s)=-\sum_{k=1}^{(n-1)/2} \arcsin{(s\bar s_k)}[\pi- \arcsin{(s\bar s_k)}]
      =-\sum_{k=1}^{(n-1)/2} [\pi^2/4-\arccos^2{(s\bar s_k)}]
\label{K_n_bos}
\end{equation}
After changing to cosine functions in the arguments, one has
\begin{equation}
\kappa_n(s) = \frac {1}{\pi^2} \, \frac {1}{n-1} \sum_{k=1}^{(n-1)/2} \left[\frac {\pi^2}{4} 
               -\arccos^2(sc_k)\right], \quad
               c_k=\cos(\frac{(2k-1)\pi}{2n}), \quad n \,\,\, odd
\label{kappa_odd_bos}
\end{equation}
In the same way, the case of even $n$ can be treated and leads to
\begin{equation}
\kappa_n(s) = \frac {1}{\pi^2} \, \frac {1}{n-1} \left\{\frac {1}{2} \left[\frac {\pi^2}{4} 
               -\arccos^2(s)\right]+\sum_{k=1}^{(n-2)/2} \left[\frac {\pi^2}{4} 
               -\arccos^2(s\bar c_k)\right] \right\}, \quad
               \bar c_k=\cos(\frac{2k\pi}{2n}), \quad n \,\,\, even
\label{kappa_even_bos}
\end{equation}
These are the analogues of the fermionic formulae (\ref{kappa_even}) and (\ref{kappa_odd}) and one
sees that basically the $\arcsin^2$ has been replaced with $\arccos^2$. The value for $s=1$
is now
\begin{equation}
\kappa_n(1) = \frac {1}{12} (1+ \frac {1}{n})
\label{kappa(1)_bos}
\end{equation}
whereas the behaviour for $s \rightarrow 0$ is linear
\begin{equation}
\kappa_n(s) = \frac {1}{2 \pi^2} \frac {1}{n-1}\, \cot(\frac {\pi}{2n})\, s \, ,
                \quad s \rightarrow 0
\label{kappa(0)_bos}
\end{equation}
The limit $n \rightarrow 1$ can be taken and gives the slope 1/4 found already in 
(\ref{kappa_bos}). Again, the cases $n=2,3$ give the simplest formulae
\begin{eqnarray}
\kappa_2(s) &=&  \frac{1}{8} \left[1-\frac {4}{\pi^2}\arccos^2(s)\right] \\
\kappa_3(s) &=&  \frac{1}{8}  \left[1-\frac {4}{\pi^2}\arccos^2(s\sqrt{3}/2) \right]
\label{kappa_23_bos}
\end{eqnarray}
%
%
%
\begin{figure}
\center
\includegraphics[scale=0.9]{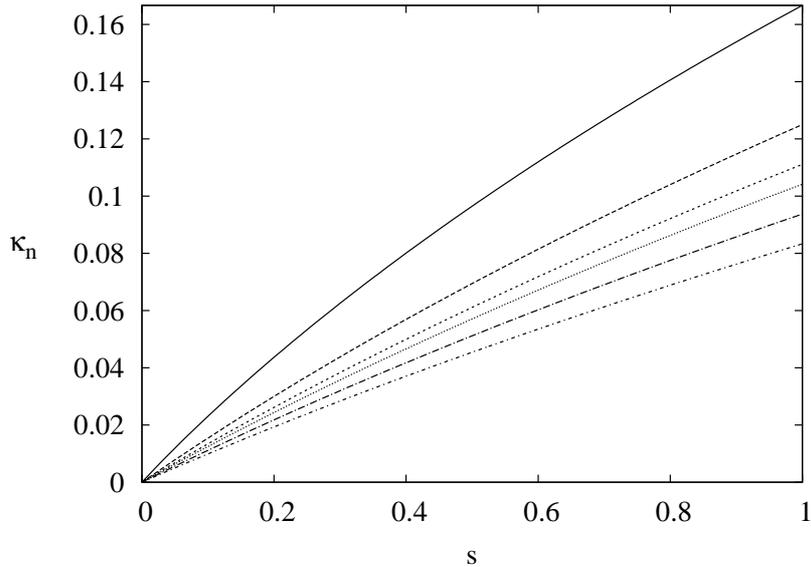}
\caption{Coefficients $\kappa_n(s)$ for the bosonic defect.
The $n$-values are, from top to bottom, $n=1,2,3,4,8,\infty.$}
\label{fig:kappa_fermion}
\end{figure}
%
%
In fig. 5, the functions $\kappa_n(s)$ are shown for six different values of $n$. They all rise
in a rather smooth way with negative curvature.  Included is also the limit $n \rightarrow \infty$, 
for which one only has to calculate the integral $I_1$.
This is  
\begin{equation}
I_1 = -\int_{0}^{\infty} \dd \varepsilon \; \left[ \ln(2 \sh \omega)- \omega \right]
\label{integral_I_1_bos}
\end{equation}
and has the derivative 
\begin{equation}
I'_1(s) = \frac {1}{s} \ln (1+s)
\label{integral_I_1_bos}
\end{equation}
which upon integration gives a dilogarithm and 
\begin{equation}
\kappa_{\infty} = -\frac {1}{\pi^2} \mathrm{Li_2}(-s)
\label{kappa_inf_bos}
\end{equation}
Thus $\kappa_{\infty}$, which describes the scaling of the largest eigenvalue of $\rho$, is a
simple, but non-elementary function, as in the fermionic case.

\section{Conclusion}

We have considered defects in critical chains of free particles which can be varied in such a 
way that one can go continuously from a homogeneous system to one cut in two pieces. 
In the RDM, this leads to a characteristic rise of the single-particle eigenvalues by which 
the entanglement across the defect becomes smaller. Both for fermions and for bosons, the
change of the eigenvalues is described by a simple functional relation. This allows to obtain
closed expressions for the entanglement entropies in the asymptotic region. In this sense, one
is dealing here with a fully soluble problem.
 
The bosonic chain, modeling a system with a conformal interface, is somewhat subtle. In order
to avoid a zero-energy mode, which would spoil the correlation function approach and does  
not contribute to the logarithmic term anyway \cite{Sakai/Satoh08}, one has to work slightly 
off-critical. The numerical extrapolations for the available sizes then do not give the same 
perfect agreement with the analytical results as for the fermionic defect. Nevertheless, they 
reproduce them over most of the parameter space. 

An interesting point is the validity of the functional relation (\ref{omega_bos}) also away
from criticality. This offers the possibility to obtain exact results also in this case,
since the $\varepsilon_k$ are explicitly known and equidistant in the infinite
system \cite{Peschel/Chung99,review09}. In the critical region, one then finds the same  
behaviour as in (\ref{kappa_def}) with $L$ replaced by the correlation length $\xi$. 
One can also see that one needs quite large values of $\xi$ to observe the exact value of
$\kappa_n(s)$ if $s$ becomes small.

Finally, one should mention that simply reducing the central spring constant in the oscillator
chain does not lead to a varying $\kappa_n$. The eigenvalues $\omega_k$ then increase in
a similar way as here and the entanglement entropy becomes smaller, but the asymptotic variation 
with $L$ remains unchanged.

\vspace{1cm}

\begin{acknowledgements}
We thank Pasquale Calabrese for stimulating correspondence on the topic and Jens Eisert for 
an interesting discussion. V.E. acknowledges financial support by the ERC grant QUERG.
\end{acknowledgements}

\appendix

\section*{Appendix: Transfer matrix in the Gaussian model}

We want to show here that the relation (\ref{omega_bos}) can also be obtained from the
transfer-matrix excitations in the associated two-dimensional model, as done in 
\cite{Eisler/Peschel10} for the fermionic case. 

Consider a Gaussian model on a square lattice with variables $\phi$, $-\infty< \phi< \infty$, and
coupling constant $K$ such that $K(\phi-\phi')^2/2$ is the energy of neighbouring sites. The 
symmetrized row transfer matrix $W=V_1^{1/2}V_2V_1^{1/2}$ consists of the contribution from vertical 
($V_1$) and horizontal ($V_2$) bonds.
These are given by
\begin{equation}
V_{1} =\exp(\,\frac {1}{2} K^* \sum_n \frac {\partial^2}{\partial \phi_n^2} )\; ,  \quad \quad
V_{2} =\exp(\,-\frac {1}{2} K \sum_n (\phi_n-\phi_{n+1})^2 )
\label{transfer1}
\end{equation}
where $K^*=1/K$. After a Fourier transformation with open boundaries and momenta $q$, they 
become 
\begin{equation}
V_{1} =\exp(\,\frac {1}{2} K^* \sum_q \frac {\partial^2}{\partial \phi_q^2} )\; , \quad \quad
V_{2} =\exp(\,-\frac {1}{2} K \sum_q \Omega_q^2 \phi_q^2 )
\label{transfer2}
\end{equation}
with $\Omega_q^2 = 2(1-\cos q)=4\sin^2(q/2)$. These can be expressed in terms of creation and
annihilation operators $b_q^{\dagger},b_q$ for oscillators with mass 1 and frequency $\Omega_q$. 
Then
\begin{equation}
V_{1} =\exp(\,\frac {1}{4} K^* \sum_q \Omega_q(b_q-b_q^{\dagger})^2)\; , \quad \quad
V_{2} =\exp(\,-\frac {1}{4} K \sum_q \Omega_q (b_q+b_q^{\dagger})^2 )
\label{transfer3}
\end{equation}
In order to obtain $W$ as a single exponential, one forms Heisenberg operators with $V_1^{1/2}$ and
$V_2$. These are
\eq{
\begin{split}
V_1^{1/2} &\left( \begin{array}{c}b_q \\ b_q^{\dag}\end{array}\right) V_1^{-1/2} =
\left(\begin{array}{cc} 1+\lambda_1 & -\lambda_1 \\ \lambda_1 & 1-\lambda_1 \end{array}\right)
\left( \begin{array}{c}b_q \\ b_q^{\dag}\end{array}\right) 
= M_1^{1/2}\left( \begin{array}{c}b_q \\ b_q^{\dag}\end{array}\right) \\
V_2 &\left( \begin{array}{c}b_q \\ b_q^{\dag}\end{array}\right) V_2^{-1} =
\left(\begin{array}{cc} 1+2\lambda_2 & 2\lambda_2 \\ -2\lambda_2 & 1-2\lambda_2
\end{array}\right)
\left( \begin{array}{c}b_q \\ b_q^{\dag}\end{array}\right)
= M_2\left( \begin{array}{c}b_q \\ b_q^{\dag}\end{array}\right)
\end{split}
\label{heisenberg1}}
with the notation $\lambda_1=K^*\Omega_q/4$ and $\lambda_2=K\Omega_q/4$. This gives for $W$
\eq{
W \left( \begin{array}{c}b_q \\ b_q^{\dag}\end{array}\right) W^{-1} =
 M_1^{1/2} M_2 M_1^{1/2}\left( \begin{array}{c}b_q \\ b_q^{\dag}\end{array}\right) = 
\left(\begin{array}{cc} \;a & \,b \\
-b & \,d \end{array}\right)
\left( \begin{array}{c}b_q \\ b_q^{\dag}\end{array}\right)
\label{heisenberg2}}
The antisymmetric matrix on the right has determinant 1 and thus eigenvalues $\ee^{\pm\gamma_q}$ 
where $\ch \gamma_q = (a+d)/2$ is half the trace. Explicitly, 
\begin{equation}
\ch \gamma_q = 1+ \frac {\Omega_q^2}{2}, \quad \mathrm{or} \quad  \sh (\gamma_q/2) = \Omega_q/2
\label{gammaq}
\end{equation}
Introducing new boson operators via
\eq{
\left( \begin{array}{c}\beta_q \\ \beta_q^{\dag}\end{array}\right) = 
\left(\begin{array}{cc}
\ch \theta_q & \sh \theta_q \\
\sh \theta_q & \ch \theta_q
\end{array}\right)
\left( \begin{array}{c} b_q \\ b_q^{\dag}\end{array}\right)=
 U \left( \begin{array}{c} b_q \\ b_q^{\dag}\end{array}\right)
\label{cantr}}
the row transfer matrix then becomes 
\begin{equation}
W = A \; \exp(\;-\sum_{q}\gamma_q \beta_q^\dag \beta_q )
\label{vdiag}
\end{equation}
The parameter in the transformation (\ref{cantr}) is given by
\begin{equation}
\exp(-4\,\theta_q) = \frac {1}{K^2}(1+ \frac {\Omega_q^2}{4})
\label{theta}
\end{equation}
and depends on the value of $K$, in contrast to $\gamma_q$.
%
%
\begin{figure}
\center
\includegraphics[scale=0.35]{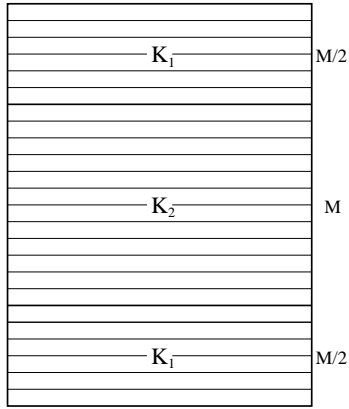}
\caption{Strip geometry defining the transfer matrix $W_{tot}$.}
\label{fig:strip}
\end{figure}
%
%

Consider now a strip as shown in fig. 6 where the three sections have coupling constants 
$K_1$, $K_2$ and $K_1$. The total transfer matrix then is $W_{tot}= W_1^{M/2} W_2^M W_1^{M/2}$
and the relation analogous to (\ref{heisenberg2}) contains the matrix
\begin{equation}
M_{tot}= U(\theta_1)^{-1}E^{M/2}U(\theta_1)U(\theta_2)^{-1}E^M U(\theta_2) U(\theta_1)^{-1}
         E^{M/2}U(\theta_1)
\label{inhom1}
\end{equation}
where $E$ is diagonal with entries $\ee^{\pm\gamma_q}$ and the index $q$ of the $\theta$
has been suppressed. But the products of the $U$'s are just $U(\theta_1-\theta_2) = U(\theta)$ where
\begin{equation}
\exp(2\,\theta) = \frac {K_1}{K_2}   
\label{theta_diff}
\end{equation}
is independent of $\Omega_q$. Using cyclic permutation, the trace is that of the matrix
\begin{equation}
  \tilde M_{tot} =  U(\theta)E^M U(\theta)^{-1}E^M
\label{inhom2}
\end{equation}
and calling the eigenvalues $\ee^{\pm 2\omega_q}$ one finds
\begin{equation}
\ch 2\omega_q = \ch ^2(M\gamma_q) + \ch(2\theta) \;\sh ^2(M\gamma_q)
\label{gammaq1}
\end{equation}
or equivalently, writing $M\gamma_q=\varepsilon_q$ and $\ch \theta =1/s$,
\begin{equation}
\sh \omega_q = \frac {1}{s} \;\sh \varepsilon_q ,
\label{gammaq2}
\end{equation}
This is the relation (\ref{omega_bos}). Due to (\ref{theta_diff}), the parameter $s$ is
defined in the same way as for the chain.

\noindent Remarks:

(i) Multiplying the symmetrized transfer matrices $W_1$ and $W_2$ leads to particular vertical 
bonds at the interface: $K_0^*=(K_1^*+K_2^*)/2$, i.e. $1/K_0=(1/K_1+1/K_2)/2$. This the choice
made in the chain calculation. 

(ii) By rescaling the variables as in section 3, one can convert the system into a homogeneous
one with two defect lines. Repeating the transfer-matrix calculation, one finds again the spectrum
(\ref{gammaq1}),(\ref{gammaq2}). 

(iii) The considerations here are not restricted to the critical point. However, the relation
to the RDM via a conformal mapping is limited to criticality.

\section*{References}

\providecommand{\newblock}{}

\end{document}